\documentclass[12pt]{article}
\usepackage{latexsym}
\usepackage{graphicx}
\thicklines                         % thick straight lines for pictures (LaTeX)
\long \def \blockcomment #1\endcomment{}
\begin{document}           % End of preamble and beginning of text.

\baselineskip=0.33333in
%\begin{quote} \raggedleft TAUP 2846-2007
%\end{quote}
%\title{A Sample Document}  % Declares the document's title.
%\author{Leslie Lamport}    % Declares the author's name.
%\date{December 12, 1984}   % Deleting this command produces today's date.
\vglue 0.5in
%\maketitle                 % Produces the title.

\begin{center}{\bf On the Quantum Mechanical State \\
of the $\Delta^{++}$ baryon}
\end{center}
\begin{center}E. Comay$^*$
\end{center}

\begin{center}
Charactell Ltd. \\
P.O. Box 39019, Tel Aviv 61390 \\
Israel
\end{center}
\vglue 0.5in
\vglue 0.5in
\noindent
%\vglue 0.2in
\vglue 0.5in
\noindent
Abstract:

The $\Delta^{++}$ and the $\Omega^-$ baryons have been used as the
original reason for the construction of the Quantum Chromodynamics
theory of Strong Interactions. The present analysis relies on the
multiconfiguration structure of states which are made of several
Dirac particles.
It is shown that this property, together with the very strong
spin-dependent interactions of quarks provide an acceptable
explanation for the states of these baryons and remove the classical
reason for the invention of color within
Quantum Chromodynamics. This explanation is
supported by several examples that show a Quantum Chromodynamics'
inconsistency with experimental results. The same arguments provide an
explanation for the problem called the proton spin crisis.

%\vglue 0.5in
%\noindent
%Key Words: proton-proton collision, very high energy, Large Hadron Collider,
%Magnetic Monopoles

\newpage
\noindent
{\bf 1. Introduction}
\vglue 0.33333in

It is well known that writing an atomic state as a sum of terms, each of
which belongs to a specific configuration is a useful tool for
calculating electronic properties of the system. This issue has already
been recognized in the early days of quantum mechanics [1]. For this purpose,
mathematical tools (called angular momentum algebra) have been developed
mainly by Wigner and Racah [2]. Actual
calculations have been carried out since the early days of electronic
computers [3]. Many specific properties of atomic states have been
proven by these calculations. In particular, these
calculations have replaced
guesses and conjectures concerning the mathematical form of atomic
states by evidence based on a solid mathematical basis.
In this work, a special emphasis is
given to the following issue: Contrary to a naive expectation, even the ground
state of a simple atom is written as a sum of more than one configuration.
Thus, the apparently quite simple closed shell
ground state of the two electron He
atom, having $J^\pi = 0^+$,
{\em disagrees} with the naive expectation where the two electrons are
just in the $1s^2$ state. Indeed, other configurations where individual
electrons take higher angular momentum states (like $1p^2,\;1d\,^2$, etc.)
have a non-negligible part of the state's description [3].
The multiconfiguration description of the ground state of the He atom
proves that shell model notation of state is far from being complete.
Notation of this model can be regarded as an anchor configuration defining
the $J^\pi$ of the state. Therefore,
all relevant configurations must have the same
parity and their single-particle angular momentum
must be coupled to the same $J$.

This paper discusses some significant elements of this scientific
knowledge and explains its crucial role in a quantum mechanical
description of the states of the $\Delta^{++}$, the $\Delta ^-$
and the $\Omega^-$ baryons. In
particular, it is proved that these baryons do not require the
introduction of new structures (like the $SU(3)$ color)
into quantum mechanics. A by-product
of this analysis is the settlement of the ``proton spin crisis" problem.

The paper is organized as follows.
The second section describes briefly some properties of angular
momentum algebra. It is proved in the third section that ordinary
laws of quantum mechanics explain why the states of the
$\Delta ^{++},\;\Delta ^-$ and $\Omega ^-$ baryons are consistent
with the Pauli exclusion principle. This outcome is used in the
fourth section for showing that QCD does not provide
the right solution for hadronic states. The problem called
``proton spin crisis" is explained in the fifth section.
The last section contains
concluding remarks.

\vglue 0.66666in
\noindent
{\bf 2. Some Features of the Electronic Angular Momentum Algebra}
\vglue 0.33333in

Consider the problem of a bound state of $N$ Dirac particles. (Baryonic
states are extremely relativistic. Therefore, a relativistic formulation
is adopted from the beginning.) This system is described as an
eigenfunction of the Hamiltonian. Thus, the time variable is removed
and one obtains a problem of $3N$ spatial variables for each of the four
components of a Dirac wave function. It is shown here how angular
momentum algebra can be used for obtaining a dramatic simplification of
the problem.

The required solution is constructed as a sum of terms, each of which
depends on all the independent variables mentioned above. Now
angular momentum is a good quantum number of a closed system and parity
is a good quantum number for systems whose state is determined by
strong or electromagnetic interactions. Thus, one takes advantage of
this fact and uses only terms that have the required angular momentum
and parity, denoted by $J^\pi$. (Later, the lower case $j^\pi$ denotes
properties of a bound spin-1/2 single particle.)

The next step is to write each term as a product of $N$
single particle Dirac functions,
each of which has a well defined angular momentum and parity. The
upper and lower parts of a Dirac function have opposite parity (see [4],
p. 53). The angular coordinates of the
two upper components of the Dirac function have
angular momentum $l$ and they are coupled with the spin
to an overall angular momentum $j=l\pm 1/2\; (j> 0)$. The two lower components
have angular momentum $(l\pm 1) \ge 0$ and together with the
spin, they are coupled to the same $j$. The spatial angular momentum
eigenfunctions having an eignevalue $l$, make a set of $(2l + 1)$
members denoted by $Y_{lm}(\theta ,\phi )$, where $\theta ,\,\phi$
denote the angular coordinates and $-l \le m \le l$ [2].

It is shown below how configurations can be used for describing
a required state whose parity and overall spin are known.

\vglue 0.66666in
\noindent
{\bf 3. The $\Delta ^{++}$ State}
\vglue 0.33333in

The purpose of this section is to describe how the state of each of the
four $\Delta $ baryons can be constructed in a way that abides by
ordinary quantum mechanics of a system of three fermions.
Since the $\Delta ^{++}$(1232) baryon has 3
valence quarks of the $u$ flavor,
the isospin $I=3/2$
of all four $\Delta $ baryons is fully symmetric. Therefore,
the space-spin components of the
3-particle terms must be antisymmetric.
(An example of relevant
nuclear states is presented at the end of this section.)
Obviously, each of the 3-particle
terms must have a total spin $J=3/2$ and an even parity. For
writing down wave functions of this kind, single particle wave
functions having a definite $j^\pi $ and appropriate radial functions
are used. A product of {\em n} specific $j^\pi $ functions is called a
configuration and the total wave function takes the form of a
sum of terms, each of which is associated with a configuration.
Here {\em n}=3 and
only even parity configurations are used. Angular momentum
algebra is applied to the single particle wave functions and
yields an overall $J=3/2$ state. In each configuration, every pair
of the $\Delta ^{++}$
{\em u} quarks must be coupled to an antisymmetric state. $r_j$ denotes
the radial coordinate of the jth quark.

Each of the A-D cases described below contains one configuration and
one or several antisymmetric 3-particle terms. The radial functions
of these examples are adapted to each case.

Notation: $f_i(r_j),\; g_i(r_j),\; h_i(r_j)$ and $v_i(r_j)$ denote radial
functions of
Dirac single particle $1/2^+,\;1/2^-,\;3/2^-$ and $3/2^+$
states, respectively. The
index $i$ denotes the excitation level of these functions. Each of these
radial functions is a two-component function, one for the upper
2-component spin and the other for the lower 2-component spin
that belong to a 4-component Dirac spinor.

\begin{itemize}

\item[{A.}] In the first example all three particles have
the same $j^\pi$,
\begin{equation}
f_0(r_0)f_1(r_1)f_2(r_2)\;1/2^+\;1/2^+\;1/2^+.
\label{eq:DPP1}
\end{equation}
Here the spin part is fully symmetric and yields a total spin of 3/2.
The spatial state is fully antisymmetric. It is obtained from the 6
permutations of the three orthogonal $f_i(r_j)$ functions divided by
$\sqrt 6$. This configuration is regarded as the anchor configuration
of the state.
\item[{B.}] A different combination of $j_i=1/2$ can be used,
\begin{equation}
f_0(r_0)g_0(r_1)g_1(r_2)\;1/2^+\;1/2^-\;1/2^-.
\label{eq:DPP2}
\end{equation}
Here, the two single particle
$1/2^-$ spin states are coupled symmetrically to j=1 and they have two
orthogonal radial functions $g_i$. The full expression can be
antisymmetrized.

\item[{C.}] In this example,
just one single particle is in an angular excitation state,
\begin{equation}
f_0(r_0)f_0(r_1)v_0(r_2)\;1/2^+\;1/2^+\;3/2^+.
\label{eq:DPP3}
\end{equation}
Here we have two $1/2^+$ single particle functions having the same
non-excited radial function. These spins are coupled
antisymmetrically to a spin zero two particle state. These
spins have the same non-excited radial function. The third
particle yields the total $J=3/2$ state. The full expression can
be antisymmetrized.

\item[{D.}] Here all the three single particle
$j^\pi $ states take different values.
Therefore, the radial functions are free to take the lowest
level,
\begin{equation}
f_0(r_0)g_0(r_1)h_0(r_2)\;1/2^+\;1/2^-\;3/2^-.
\label{eq:DPP4}
\end{equation}
Due to the different single particle spins, the antisymmetrization
task
of the spin coordinates can easily be done. (The spins can be coupled
to a total $J=3/2$ state in two different ways. Hence, two different
terms belong to this configuration.)

\end{itemize}

The examples A-D show how a Hilbert space basis for the $J^\pi=3/2^+$
state can be constructed for three identical fermions.
Obviously, more configurations can be added
and the problem can be solved by ordinary
spectroscopic methods. It should
be noted that unlike atomic states,
the very strong spin dependent interactions of hadrons
are expected to yield a higher configuration mixture.

\begin{table} [b]
\caption {Nuclear A=31 Isospin State Energy levels (MeV)}
\vglue 0.1in
\centering
\begin{tabular} {| c | c | c | c | c | c |}
% & & & \\
\hline
% \\[1pt]
$J^\pi$ & I (T)$^a$ & $^{31}$Si & $^{31}$P & $^{31}$S & $^{31}$Cl$^b$ \\
% \\[1pt]
\hline
% \\[1pt]
 & & & & & \\
$1/2^+$ & $1/2$ & - & 0 & 0 & - \\
 & & & & & \\
$3/2^+$ & $3/2$ & 0 & 6.38 & 6.27 & 0 \\
 & & & & & \\
$1/2^+$ & $3/2$ & 0.75 & 7.14 & 7.00 & - \\
 & & & & & \\
\hline
\multicolumn{6}{|l|}
{$^a$ I,T denote isospin in particle physics} \\
\multicolumn{6}{|l|}
{$\;\;\;$and nuclear physics, respectively.} \\
\multicolumn{6}{|l|}
{$^b$ The $^{31}$Cl data is taken from [6].} \\
\hline
\end{tabular}
\end{table}

It is interesting to note that a similar situation is found in
nuclear physics.
Like the {\em u,d} quarks,
the proton and the neutron are spin-1/2 fermions belonging to an
isospin doublet. This is the basis for the common symmetry
of isospin states described below.
Table 1 shows energy levels of each of the
four A=31 nuclei examined [5].
Each of these nuclei has 14
protons and 14 neutrons that occupy a set of inner closed shells and
three nucleons outside these shells. (The closed shells are $1/2^+,\;
3/2^-,\; 1/2^-,$ and $5/2^+$. The next shells are the $1/2^+$ that
can take 2 nucleons of each type and the $3/2^+$ shell that is higher than
the $1/2^+$ shell. See [7], p. 245.) The state is characterized by these
three nucleons that define the values of total spin, parity and
isospin. The first line of table 1 contains data of the ground
states of the $I=1/2$ $^{31}$P and $^{31}$S nuclei.
The second line contains
data of the lowest level of the $I=3/2$ state of the four nuclei. The
quite small energy difference between the $^{31}$P and $^{31}$S
excited states illustrates
the quite good
relative accuracy of the isospin approximation. The third line of
the table shows the first excited $I=3/2$ state of each of
the four nuclei. The gap between
states of the third and the second lines is nearly, but not precisely,
the same. It provides another example of
the relative goodness of the isospin approximation.

The nuclear states described in the first and the second lines of
table 1 are relevant to
the nucleons and the $\Delta $ baryons of particle
physics. Indeed,
the states of both systems are characterized by three fermions that may belong
to two different kinds and where isospin is a useful quantum number.
Here the neutron and the proton
correspond to the ground state of $^{31}$P and $^{31}$S, respectively,
whereas energy states of the second line of the table correspond to
the four $\Delta $ baryons. Every nuclear energy state of table 1 has a
corresponding baryon that has the same spin, parity, isospin
and the $I_z$ isospin component. Obviously, the dynamics of the
nuclear energy levels is completely different from hadronic
dynamics. Indeed, the nucleons are composite spin-1/2 particles whose
state is determined by the strong nuclear force. This is a residual
force characterized by a rapidly decaying attractive force at the
vicinity of the nucleon size and a strong repulsive force at a smaller
distance (see [7], p. 97). On the other hand, the baryonic quarks are
elementary pointlike spin-1/2 particles whose dynamics differs
completely from that of the strong nuclear force. However,
both systems are made of fermions and the
spin, parity and isospin analogy demonstrates that
{\em the two systems have the same internal symmetry.}

The following property of the second line
of table 1 is interesting and important. Thus, all nuclear states
of this line have the same {\em symmetric}
$I=3/2$ state. Hence, due to the Pauli exclusion principle, all of them have
the same {\em antisymmetric space-spin state.}
Now, for the $^{31}$P and $^{31}$S nuclei, this
state is an excited state because they have lower states having
$I=1/2$. However, the $^{31}$Si and  $^{31}$Cl nuclei have no $I=1/2$ state,
because their $I_z=3/2$. Hence, the second line of table 1 shows
{\em the ground state of the $I_z=3/2$ nuclei}. It will be shown
later that this conclusion is crucial for having a good
understanding of an analogous quark system. Therefore it is called
{\em Conclusion A}.

Now, the $^{31}$Si has three neutrons outside the 28 nucleon closed
shells and the $^{31}$Cl has three protons outside these shells.
Hence, the outer shell of
these two nuclear states consists of three identical fermions
which make the required ground state. Relying on this nuclear physics example,
one deduces that {\em the Pauli exclusion principle is completely
consistent with three identical fermions in a $J^\pi = 3/2^+$
and $I=3/2$ ground state.} The data of table 1
are well known in nuclear physics.

A last remark should be made before the end of this section.
As explained in the next section, everything
said above on the isospin quartet
$J^\pi = 3/2^+$ states of the three {\em ud} quark flavor
that make the four $\Delta $
baryons, holds for the full decuplet of the three
{\em uds} quarks, where, for
example, the $\Omega ^-$ state is determined by the three {\em sss} quarks.
In particular, like the $\Delta ^{++}$ and the $\Delta ^-$,
{\em the $\Omega ^-$ baryon
is the ground state of the three sss quarks and each of the baryons of
the decuplet has an antisymmetric space-spin wave function.}

\vglue 0.66666in
\noindent
{\bf 4. Discussion}
\vglue 0.33333in

It is mentioned above that spin-dependent interactions are much stronger
in hadronic states than in electronic states. This point is illustrated
by a comparison of the singlet and triplet states of the positronium [8]
with the $\pi ^0$ and $\rho ^0$ mesons [9].
The data are given in table 2. The fourth column of the table shows
energy difference between each of the $J^\pi =1^-$ states and the
corresponding $J^\pi =0^-$ state. The
last column shows the ratio between this difference and the energy of the
$J^\pi =0^-$ state.

\begin{table} [b]
\caption {Positronium and meson energy}
\vglue 0.1in
\centering
\begin{tabular} {| c | c | c | c | c |}
% & & & \\
\hline
% \\[1pt]
Particles & $J^\pi$ & Mass & M($1^-$) - M($0^-$) & $\Delta $M/M($0^-$) \\
% \\[1pt]
\hline
% \\[1pt]
 & & & & \\
Positronium & $0^-$ & $\sim 1022$ KeV  & - & -\\
 & & & & \\
Positronium & $1^-$ & $\sim 1022$ KeV  & 0.00084 eV & 8.2 10$^{-10}$ \\
  & & & & \\
\hline
  & & & & \\
$\pi ^0$ & $0^-$ & 135 MeV  & - & -\\
 & & & & \\
$\rho ^0$ & $1^-$ & 775 MeV  & 640 MeV & 4.7 \\
 & & & & \\
\hline
\end{tabular}
\end{table}

Both electrons and quarks are spin-1/2 pointlike particles.
The values of the last column
demonstrate a clear difference between electrically charged
electrons and quarks that participate in strong interactions.
Indeed, the split between the two electronic states
is {\em very small}. This is
the reason for the notation of {\em fine structure} for the
spin dependent split between electronic states of the same
excitation level (see [10], p.225).
Table 2 shows that the corresponding
situation in quark systems is larger by more than 9 orders of magnitude.
Hence, spin dependent interactions in hadrons are very strong
and make an important contribution to the state's energy.

Now, electronic systems in atoms satisfy the Hund's rules (see
[10], p. 226). This rule says that in a configuration, the state
having the highest spin is bound stronger. Using this rule {\em and}
the very strong spin-dependent hadronic interaction which is
demonstrated in the last column of table 2, one concludes that the
anchor configuration A of the previous section really describes a very
strongly bound state of the $\Delta ^{++}$ baryon. In particular,
the isospin doublet $J^\pi =1/2^+$ state of the neutron and the proton
correspond to the same $J^\pi =1/2^+$ of the ground state of the $A=31$
nuclei displayed in the first line of table 1. The isospin
quartet of the $\Delta $ baryons correspond to the isospin
quartet of the four nuclear states displayed in the second line
of table 1.

Here the significance of Conclusion A of the previous section becomes clear.
Indeed, an analogy is found between the two nuclear states of the
$I=3/2$ and $I_z=\pm 1/2$, namely the $^{31}$P and the $^{31}$S
are {\em excited states} of these nuclei
whereas the $I=3/2$ and $I_z=\pm 3/2$,
namely the $^{31}$Si and the $^{31}$Cl states are the {\em ground states}
of these nuclei. The same pattern is found in the particle physics
analogue. The $\Delta ^+$ and the $\Delta ^0$ are {\em excited states}
of the proton and the neutron, respectively. This statement relies
on the fact that both the proton and the $\Delta ^+$ states
are determined by the {\em uud} quarks. Similarly, the neutron
and the $\Delta ^0$ states are determined by the {\em udd} quarks.
On the other hand, in the case of the $^{31}$Si and the $^{31}$Cl nuclei,
the $I=3/2$ and $J^\pi =3/2^+$ states are the {\em ground states}
of these nuclei. The same property holds for the
$\Delta ^{++}$ and the $\Delta ^-$, which are the {\em ground states}
of the {\em uuu} and {\em ddd} quark systems, respectively.

The relationship between members of the lowest energy $J^\pi=1/2^+$
baryonic octet and members of the $J^\pi=3/2^+$ baryonic decuplet
can be described as follows. There are 7 members of the
decuplet that are excited states of corresponding members of
the octet. Members of each pair are made of the same specific
combination of the {\em uds} quarks. The $\Delta ^{++},\;\Delta^-$
and $\Omega ^-$ baryons have no counterpart in the octet. Thus, in
spite of being a part of the decuplet whose members
have space-spin antisymmetric states, these
three baryons are the {\em ground state} of the {\em uuu, ddd} and
{\em sss} quarks, respectively.

This discussion can be concluded by the following statements: {\em
The well known laws of quantum
mechanics of identical fermions provide an interpretation of
the $\Delta ^{++}, \;\Delta ^-$ and $\Omega ^-$ baryons,
whose state is characterized by three
uuu, ddd and sss quarks, respectively.
There is no need for any fundamental change in physics in
general and for the invention of color in particular. Like all
members of the decuplet, the states of these baryons abide by
the Pauli exclusion principle.}
Hence, one wonders why particle physics textbooks regard precisely
the same situation of the four $\Delta $ baryons
as a fiasco of the Fermi-Dirac statistics (see [11], p. 5).

The historic reasons for the QCD creation are the states of the
$\Delta ^{++}$ and the $\Omega ^{-}$ baryons. These baryons,
each of which has three quarks of the same flavor,
are regarded as the {\em classical reason} for the QCD invention
(see [12], p. 338). The analysis presented above shows that this
argument does not hold water. For this reason, one wonders whether
QCD is really a correct theory. This point is supported by the
following examples which show that QCD
is inconsistent with experimental results.

\begin{itemize}
\item[{1.}] The interaction of hard real photons with a proton
is practically the same as its interaction with a neutron [13]. This
effect cannot be explained by the photon interaction with the nucleons'
charge constituents, because these constituents
take different values for the proton and
the neutron. The attempt to recruit Vector Meson Dominance (VMD)
for providing an explanation is unacceptable. Indeed, Wigner's analysis
of the irreducible representations of the Poincare group [14,15] proves
that VMD, which mixes a massive meson with a massless photon, is
incompatible with Special relativity. Other reasons of this kind have
also been published [16].
\item[{2.}] QCD experts have predicted the existence of
strongly bound pentaquarks [17,18]. In spite of a long search, the
existence of pentaquarks has not been confirmed [19].
By contrast, correct physical ideas used for predicting genuine particles,
like the positron and the $\Omega ^-$, have been validated by experiment
after very few years (and with a technology which is very very poor
with respect to that used in contemporary facilities).

\item[{3.}] QCD experts have predicted the existence of chunks of
Strange Quark Matter (SQM) [20]. In spite of a long search, the
existence of SQM has not been confirmed [21].
\item[{4.}] QCD experts have predicted the existence of glueballs [22].
In spite of a long search, the
existence of glueballs has not been confirmed [9].
\item[{5.}] For a very high energy, the proton-proton ({\em pp})
total and elastic cross section increase with collision energy [9]
and the ratio of the elastic cross section to the total cross
section is nearly a constant which equals about 1/6.
This relationship between the {\em pp} cross sections
is completely different from the high energy electron-proton
({\em ep}) scattering
data where the total cross section decreases for an increasing
collision energy and the elastic cross section's portion becomes
negligible [23].
This effect proves that the proton contains a quite solid
component that can take the heavy blow of the high energy collision
and keep the entire proton intact. This object cannot be a quark,
because in energetic {\em ep} scattering, the electron strikes a
single quark and the relative part of elastic events is negligible.
QCD has no explanation for the {\em pp} cross section data [24].

\end{itemize}

\vglue 0.66666in
\noindent
{\bf 5. The Proton Spin Crisis}
\vglue 0.33333in

Another problem which is settled by the physical evidence described
above is called {\em the proton spin crisis} [25,26]. Here polarized
muons have been scattered by polarized protons. The results prove
that the instantaneous quark spin sums up to a very small
portion of the entire proton's spin. This
outcome, which has been regarded as a surprise [26],
was later supported by other experiments. The
following lines contain a straightforward explanation of this
phenomenon.

The four configurations A-D that are a part of
the Hilbert space of the $\Delta ^{++}$ baryon are used as an
illustration of the problem. Thus, in configuration A, all single
particle spins are parallel to the overall spin. The situation in
configuration B is different. Here the single particle function
$j^\pi = 1/2^-$ is a coupling of $l=1$ and $s=1/2$. This
single particle coupling has two terms whose numerical values are
called Clebsh-Gordan coefficients [2]. In one of the coupling terms,
the spin is parallel to the overall single particle angular
momentum and in the other term it is antiparallel to it. This is
an example of an internal partial cancellation of the
contribution of the single
particle spin to the overall angular momentum. (As a matter of fact,
this argument also holds for the lower pair of components of each
of Dirac spinor of configuration A. Here the lower pair of the
high relativistic system is quite large and it is made of
$l=1\;s=1/2$ coupled to $J=1/2$.)
In configuration C
one finds the same effect. Spins of the first and the second
particles are coupled to
$j_{01}=0$ and cancel each other. In the third particle the $l=2$
spatial angular momentum is coupled with the spin to $j=3/2$. Here
one also finds two terms and the contribution of their
single-particle spin-1/2
partially cancels. The same conclusion is obtained from an
analogous analysis of the spins of configuration D.

It should be pointed out that the very large hadronic
spin-dependent interaction which is demonstrated by the data of
table 2, proves that in hadronic states, one needs many
configurations in order to construct a useful basis for the Hilbert
space of a baryon. It follows that in hadrons the
internal single particle cancellation seen in configurations
of the previous section, is expected to be quite large.

Obviously,
the configuration structure of the proton relies on the same
principles. Here too, many configurations, each of which
has the quantum numbers $J^\pi = 1/2^+$, are needed for the
Hilbert space. Thus, a large
single particle spin cancellation is obtained.
Therefore, the result of [25] is obvious. It should make neither a crisis
nor a surprise.

On top of what is said above, the following argument indicates that
the situation is more complicated and the number
of meaningful configurations is even larger. Indeed, it has been
shown that beside the three valence quarks,
the proton contains additional quark-antiquark pair(s)
whose probability is about 1/2 pair [23]. It is
very reasonable to assume that all baryons share this property.
The additional quark-antiquark pair(s) increase the number of useful
configurations and of their effect on the $\Delta ^{++}$ ground
state and on the proton spin as well.

\vglue 0.66666in
\noindent
{\bf 6. Concluding Remarks}
\vglue 0.33333in

Relying on the analysis of the apparently quite simple ground state
of the He atomic structure [3], it is argued
here that {\em many} configurations are needed for describing a
quantum mechanical state of more than one Dirac particle. This effect
is much stronger in baryons, where, as shown in table 2,
spin-dependent strong interactions are very strong indeed. This effect
and the multiconfiguration
basis of hadronic states do explain the isospin quartet of the
$J=3/2^+$ $\Delta $
baryons. Here the $\Delta ^0$ and the $\Delta ^+$ are {\em excited states}
of the neutron and the proton, respectively whereas their isospin
counterparts, the $\Delta ^{++}$ and the $\Delta ^-$ are {\em ground states}
of the {\em uuu} and the {\em ddd} quark systems, respectively.
Analogous conclusions hold for all members of the $J=3/2^+$
baryonic decuplet that includes the $\Omega ^-$ baryon.
It is also shown that states of four $A=31$ nuclei are analogous to
the nucleons and the $\Delta $s isospin quartet (see table 1).

The discussion presented above shows that there is no need for
introducing a new degree of freedom (like color) in order to settle
the states of $\Delta ^{++},\;\Delta ^-$ and $\Omega ^-$ baryons with the
Pauli exclusion principle. Hence, there is no reason for the QCD
invention. Several inconsistencies of QCD with experimental data
support this conclusion.

Another aspect of recognizing implications of
the multiconfiguration structure of
hadrons is that the proton spin crisis
experiment, which shows that instantaneous spins of
quarks make a little contribution to the proton's spin [25],
creates neither a surprise nor a crisis.

%  ??????????????????????????????

\newpage
References:
\begin{itemize}
\item[{*}] Email: elicomay@post.tau.ac.il  \\
\hspace{0.5cm}
           Internet site: http://www.tau.ac.il/$\sim $elicomay

\item[{[1]}] H. A. Bethe, {\em Intermediate Quantum Mechanics} (Benjamin,
New York, 1964). (see p. 109).
\item[{[2]}] A. de-Shalit and I. Talmi, {\em Nuclear Shell Theory}
(Academic, New York, 1963).
\item[{[3]}] A. W. Weiss, Phys. Rev. 122, 1826 (1961)
\item[{[4]}] J. D. Bjorken
and S. D. Drell {\em Relativistic Quantum Mechanics} (McGraw, New York, 1964).

\item[{[5]}] P. M. Endt, Nuc. Phys. {\bf A521}, 1 (1990). (See p. 357.)

% Isospin quadruplet of A=31 nuclei.

\item[{[6]}] A. Kankainen et al., Eur. Phys. J. A 27, 67 (2006)

% Ground state of 31Cl is j^\pi=3/2^+ I=3/2

\item[{[7]}] S. S. M. Wong, {\em Nuclear Physics} (Wiley, New York, 1998).
\item[{[8]}] S. Berko and H. N. Pendleton, Ann. Rev. Nuc. Part. Sci.
{\bf 30}, 543 (1980).
\item[{[9]}] K. Nakamura et al. (Particle Data Group), J. Phys.
{\bf G37}, 075021 (2010) (URL: http://pdg.lbl.gov)
\item[{[10]}] L. D. Landau and E. M. Lifshitz, {\em Quantum Mechanics}
(Pergamon, London, 1959).
\item[{[11]}] F. Halzen and A. D. Martin, {\em Quarks and Leptons}
(Wiley, New York, 1984).
\item[{[12]}] F. E. Close, {\em An Introduction to quarks and Partons}
(Academic, London, 1979).
\item[{[13]}] T. H. Bauer, R. D. Spital, D. R. Yennie and
F. M. Pipkin {\em Rev. Mod. Phys.} {\bf 50}, 261 (1978).
\item[{[14]}] E. P. Wigner, Annals of Math., {\bf 40}, 149 (1939).
\item[{[15]}] S. S. Schweber, {\em An Introduction to Relativistic
Quantum Field Theory}, (Harper \& Row, New York, 1964). Pp. 44-53.
\item[{[16]}] E. Comay, Apeiron {\bf 10}, No. 2, 87 (2003).
\item[{[17]}] C. Gignoux, B. Silvestre-Brac and J. M. Richard,
Phys. Lett. {\bf 193}, 323 (1987).
\item[{[18]}] H. J. Lipkin, Phys. Lett. {\bf 195}, 484 (1987).
\item[{[19]}] C. G. Whol (in the 2009 report of PDG).
\item[{[20]}] E. Witten, Phys. Rev. {\bf D30}, 272 (1984).
\item[{[21]}] K. Han et al., Phys. Rev. Lett. 103, 092302 (2009).
\item[{[22]}] H. Frauenfelder and E. M. Henley, {\em Subatomic Physics},
(Prentice Hall, Englewood Cliffs, 1991).
\item[{[23]}] D. H. Perkins {\em Introductions to High Energy Physics}
(Addison-Wesley, Menlo Park, 1987) 3rd edn.
\item[{[24]}] A. A. Arkhipov,
\nolinebreak
{http://arxiv.org/PS\_cache/hep-ph/pdf/9911/9911533v2.pdf}
\item[{[25]}] J. Ashman et al. (EMC), Phys. Lett. {\bf B206}, 364 (1988).
\item[{[26]}] F. Myhrer and A. W. Thomas, J. Phys. {\bf G37}, 023101
(2010).

\end{itemize}

\end{document}